# Reststrahlen band and optical bandgaps in semiconducting CrN films

Duc V. Dinh,* Xiang Lü, and Oliver Brandt

*Paul-Drude-Institut für Festkörperelektronik, Leibniz-Institut im Forschungsverbund Berlin e.V., Hausvogteiplatz 5–7, 10117 Berlin, Germany.*

Dilara Sen, Olivia Fairlamb, and Frank Peiris

*Department of Physics, Kenyon College, Gambier, 43022 Ohio, United States*

Farihatun Lima, Alexander Bordovalos, Suresh Chaulagain, Ambalanath Shan, and Nikolas J. Podraza

*Department of Physics, University of Toledo, Toledo, 43606 Ohio, United States*

We present a comprehensive optical characterization of 200-nm-thick CrN(111) films grown simultaneously on $Al_2O_3$(0001) and AlN/$Al_2O_3$(0001) using plasma-assisted molecular beam epitaxy. Spectroscopic ellipsometry, spanning the far-infrared to ultraviolet range (0.04 – 5.5 eV), is conducted at room temperature to determine the optical constants $n$ and $k$ of the films. Spectral fits reveal two interband transitions at approximately 0.35 and 0.60 eV. In the infrared range, the ellipsometry data also reveals a pronounced Reststrahlen band stemming from transversal and longitudinal optical phonons at approximately 403 and 629 $cm^{-1}$, respectively. The relative static and high-frequency permittivities are estimated to be about 39 and 15, respectively. A Born effective charge of approximately 2.7, extracted from the far-infrared region, indicates that CrN is partially ionic.

CrN, a transition-metal nitride that crystallizes in the rock-salt structure, has attracted broad attention not only for its well-established mechanical properties — such as high hardness, wear resistance, and excellent corrosion stability [1, 2] — but also for its intriguing electronic [3–8], magnetic [9–13], and thermoelectric characteristics [14–18]. These multifunctional properties make CrN an attractive platform for integrated electronics and energy conversion technologies, particularly in applications requiring simultaneous mechanical durability, magnetic functionality, and tunable transport behavior.

Despite extensive research, the fundamental electronic structure of CrN remains a subject of active debate, with reported values spanning from 0.02 to 0.7 eV depending on measurement techniques and material quality [3, 4, 6–8, 11, 19, 20]. Temperature-dependent resistivity measurements up to 400 K typically yielded lower bandgap values, ranging from 0.02 to 0.09 eV [3, 4, 6–8, 11], while optical techniques reported larger values between 0.2 and 0.7 eV [5, 19, 20]. An exceptionally comprehensive study has estimated an indirect fundamental bandgap of (0.19 ± 0.46) eV [5]. Other measurements reveal absorption onsets around 0.64 – 0.7 eV [19, 20], which are considered to correspond to higher-energy direct transitions rather than the fundamental gap. More recent studies support the attribution of these features to direct interband transitions [5, 21]. In summary, these findings underscore both the complexity of the electronic structure of CrN and the strong sensitivity of reported bandgap values to synthesis parameters and characterization methods.

Several groups have also employed first-principles density functional theory (DFT) calculations to investigate the bandgap of CrN [13, 21–26]. These studies reveal that the predicted bandgap varies widely—from negative values [26] to as much as 2 eV [23]—depending on the exchange-correlation functional, magnetic ordering, structural distortions, strain relaxation, and other computational parameters. This theoretical variability, combined with inconsistencies in experimental measurements, underline the challenge of reliably determining the fundamental bandgap of CrN.

Recent advances in device engineering have placed phonon dynamics and infrared optical properties at the forefront of materials research. In CrN, infrared-active phonon modes and the associated Reststrahlen band have been identified through reflectance and transmittance measurements on sputtered thin films [5]. However, direct observation of these features via spectroscopic ellipsometry (SE) in epitaxial layers remains unreported. SE offers distinct advantages by simultaneously accessing both the real and imaginary components of the dielectric function, without requiring baseline corrections or Kramers-Kronig transformations typically needed for standalone transmittance or reflectance data. Despite these advantages, the optical properties of CrN have been scarcely explored using ellipsometry, with only limited prior studies available [27]. Related rock-salt transition-metal nitrides such as ScN have exhibited a pronounced Reststrahlen band [28], underscoring the role of lattice phonons in this material class and motivating further optical investigations of CrN.

Understanding the fundamental bandgap of CrN and its optical response is therefore essential to unlocking its potential across a broad spectrum of applications, including electronic, optoelectronic, and energy-related technologies. In addition, the presence of a Reststrahlen band offers opportunities for mid-infrared photonics, thermal management, and phonon-polariton-based device architectures [29, 30].

We have recently investigated the electrical properties of CrN(111) and CrN(113) films, observing two electrical bandgaps of approximately 0.15 and 0.5 eV in the intrinsic region (300 − 920 K) for both orientations [31]. In this work, we determine the dielectric function of epitaxial CrN thin films grown simultaneously on $Al_2O_3(0001)$ and $AlN/Al_2O_3(0001)$ by plasma-assisted molecular beam epitaxy (PAMBE) using SE over a broad spectral range of 0.04 – 5.5 eV. The data are analyzed using a parametric oscillator model, revealing two direct interband transitions at approximately 0.35 and 0.60 eV. A pronounced Reststrahlen band, arising from transversal and longitudinal optical phonons near 403 and 629 $cm^{-1}$, is also identified. These features are consistently observed for both substrates, confirming their intrinsic origin. Beyond this spectral analysis, we further investigate the dielectric constants, Born effective charges of CrN, providing insight into the intricate interplay between its lattice dynamics and carrier transport properties.

CrN films are grown simultaneously on singular $Al_2O_3(0001)$ and vicinal (4°-miscut) $AlN/Al_2O_3(0001)$ by PAMBE. Prior to being introduced into the ultrahigh vacuum environment, the substrates undergo a cleaning procedure involving HCl solution to remove surface oxides and contaminants. This is followed by rinsing with de-ionized water and drying with a nitrogen gun. Afterwards, the substrates were outgassed for two hours at 500 °C in a load-lock chamber attached to the MBE system. The MBE growth chamber is equipped with a high-temperature effusion cell to provide Cr (99.995 % pure). A Veeco UNI-Bulb radio-frequency plasma source is used for the supply of active nitrogen ($N^*$). The $N^*$ flux is calculated from the thickness of a GaN layer grown under Ga-rich conditions, and thus with a growth rate limited by the $N^*$ flux [32]. To avoid the formation of the trigonal $Cr_2N$ phase, the CrN films are grown at 500 °C under $N^*$-rich conditions. The structural properties of the films are characterized using a triple-axis high-resolution x-ray diffractometry (HRXRD) system (Panalytical X'Pert PRO MRD) equipped with a two-bounce hybrid monochromator Ge(220) for the $CuK_{\alpha 1}$ source ($\lambda$ = 1.540 598 Å). The thickness of the films of $\approx$ 200 nm is obtained from x-ray thickness interference fringes (see Fig. S1 in the supplementary material). Both layers are relaxed with a lattice constant of (4.145±0.001) Å. Information about the calculation of their lattice constant is provided in Fig. S2 in the supplementary material. The surface morphology of the films is imaged by atomic force microscopy in contact mode (Dimension Edge, Bruker). To investigate the electrical properties of the film, Hall-effect measurements were performed at room temperature using the van der Pauw configuration. The films exhibit a carrier concentration $n_e$ in the range of $(1.3 - 1.5) \times 10^{19}$ $cm^{-3}$ and a mobility in the range of 10 – 13 $cm^2V^{-1}s^{-1}$. SE measurements are carried out at room temperature using two Woollam ellipsometers, covering a combined spectral range from the far-infrared to the ultraviolet (0.04 – 5.5 eV). Data were collected at three angles of incidence, $\theta$ = 65, 70 and 75° (with respect to the surface normal). The optical constants of the CrN layers are extracted by fitting the SE data using a parametric multilayer model incorporating

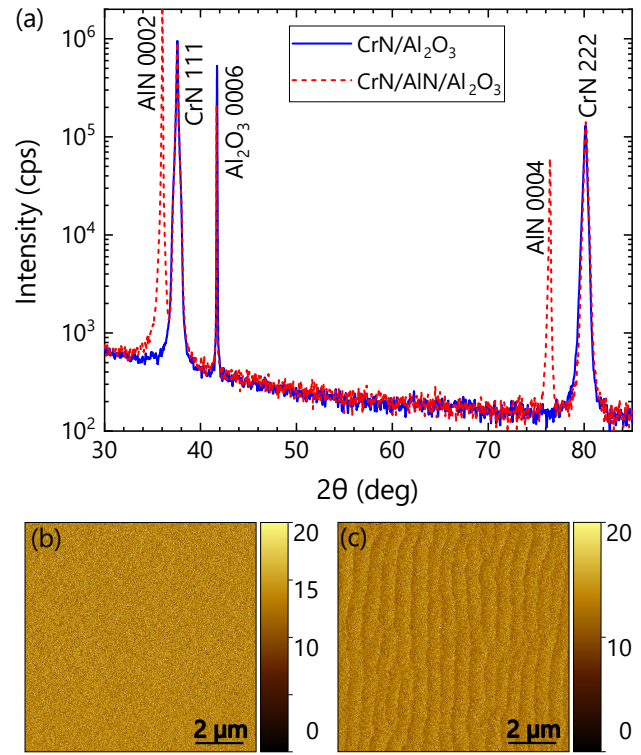

FIG. 1. (a) XRD scans of the 200-nm-thick CrN films grown simultaneously on $Al_2O_3(0001)$ and $AlN/Al_2O_3(0001)$. $10 \times 10$ $\mu m^2$ atomic force topographs of the films on (b) $Al_2O_3$ and (c) $AlN/Al_2O_3$, with the z-scale bar in nanometers. The root-mean-square roughness values of the films are 2.5 and 2.1 nm, respectively.

both electronic and phononic contributions, represented as a collection of oscillators.

Figure 1(a) shows symmetric 2θ–ω XRD scans of the CrN films grown on the two different substrate types. Both films exhibit the (111) orientation of rock-salt structured CrN, with no detectable secondary phase of trigonal $Cr_2N$. The film grown on $AlN/Al_2O_3$ exhibits a full-width at half maximum of the CrN111 x-ray rocking curve of 0.06°, compared to 0.1° for the film grown directly on $Al_2O_3$. While the magnitude of the lattice mismatch is only marginally different (−5.8% for CrN/AlN vs. 6.7% for $CrN/Al_2O_3$), the films are initially in opposite strain states, until they both fully relax. Furthermore, the different interface chemistry may lead to a higher density of structural defects for the latter film, resulting in its larger mosaicity.

Both films show relatively smooth surfaces, with root-mean-square roughness values between 2 and 2.5 nm measured over a $10 \times 10$ $\mu m^2$ area [Figs. 1(b)–1(c)].

To fit the experimental SE spectra, we construct multilayer optical models for both sample types: $CrN/Al_2O_3$ and $CrN/AlN/Al_2O_3$. Each model consists of an $Al_2O_3$ substrate, a CrN film, and a top surface roughness layer represented using the Bruggeman effective medium approximation [33]. For the film on $AlN/Al_2O_3$, an additional AlN layer is included between the substrate and the CrN film. The dielectric response of CrN is modeled using a wavelength-by-wavelength fit in the whole spectral range [34]. Model parameters — including film thicknesses and roughness parameters — are varied to achieve the best fit to the experimental Ψ and Δ spectra.

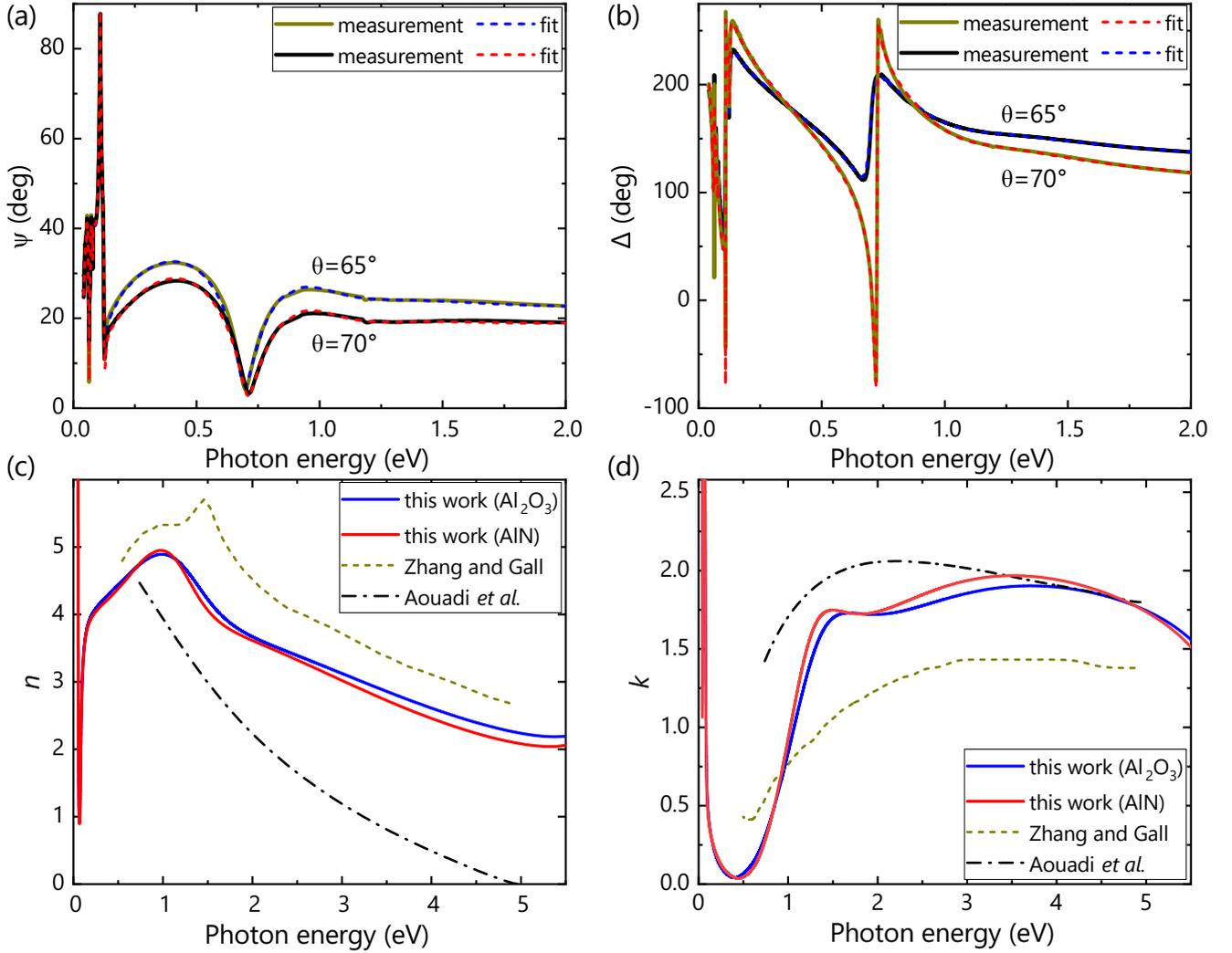

FIG. 2. (a) Ψ and (b) Δ spectra and their fits of the CrN/Al$_2$O$_3$ film. (c) Refractive index ($n$) and (d) extinction coefficient ($k$) of the films on Al$_2$O$_3$ and AlN/Al$_2$O$_3$, as obtained from the fits. Data from Zhang and Gall [5] and Aouadi et al. [27] are shown in (c) and (d) for comparison. Data from Zhang and Gall [5] cover a similar spectral range as our measurements, but only the 0.5 – 5 eV portion is shown here for clarity.

Figures 2(a)–2(b) display the measured Ψ and Δ spectra for the CrN/Al$_2$O$_3$ film, together with their corresponding fits across the 0.04 − 2.0 eV range. The complete dataset and fits over the full spectral range are shown in Fig. S3 in the supplementary material. The satisfactory fits enable the extraction of optical constants across the full spectral range of CrN, as illustrated in Figs. 2(c)–2(d). Both films exhibit very similar dispersion characteristics with only minor deviations at higher energies.

For comparison, the optical constants of sputtered films reported by Zhang and Gall [5] — derived from reflectance and transmittance measurements, and by Aouadi et al. [27] — extracted via SE, are presented in Figs. 2(c)–2(d). While Zhang and Gall [5] report higher $n$ compared to our data, their $k$ values are lower and more broadly distributed. This contrast suggests that their samples exhibit stronger reflective behavior but weaker intraband absorption, possibly due to the significantly higher electron densities in their films (in the range of $10^{20}$ cm$^{-3}$) reported in a subsequent study by the same authors[6]. Data from Aouadi et al. [27] show no indication of oscillators in either $n$ or $k$, likely due to lower material quality. In comparison, our SE measurements reveal pronounced peaks in both $n$ and $k$, indicative of more distinct electronic resonances in our epitaxial CrN films.

To analyze the electronic transitions in CrN, we fitted the imaginary part $\varepsilon_2$ of the dielectric function of both films using a combination of five oscillators: one Lorentz, one Drude and three Tauc-Lorentz (TL) oscillators [Fig. 3(a)].

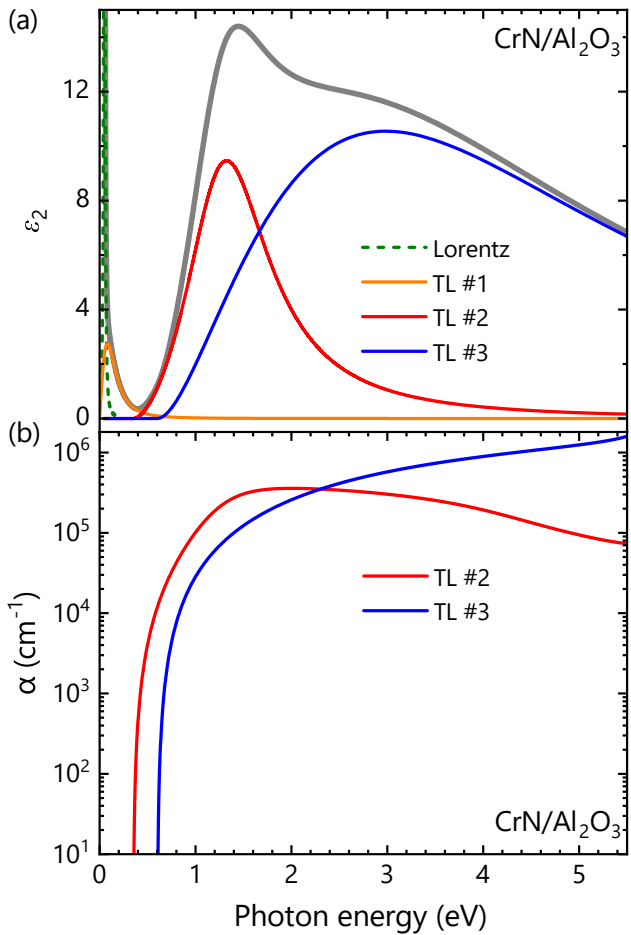

FIG. 3. (a) Lorentz and Tauc-Lorentz (TL) fitting of the imaginary part $\varepsilon_2$ of the dielectric function for the CrN/Al$_2$O$_3$ film. The thick line shows the total fit. (b) Absorption coefficients (semi-log scale) calculated for TL #2 and TL #3.

The Lorentz oscillator accounts for the strong phonon absorption observed in the far-infrared region, which will be discussed in detail later.

Among the three TL oscillators, the first one (TL #1) has a very small amplitude and has an onset of absorption near zero, and thus primarily serves as a background. Because a zero-gap oscillator cannot represent a conventional electronic interband transition, it does not arise from an electronic excitation. The second (TL #2) and the third TL (TL #3) oscillators have pronounced amplitudes with onset energies at 0.35 and 0.60 eV, respectively. These two TL oscillators, therefore, signal two associated interband transitions. The higher energy transition is also observed as an absorption onset at 0.64 eV in Fourier transform infrared spectroscopy measurements at room temperature (see Fig. S4 in the supplementary material). While transmittance alone is not fully quantitative without corresponding reflectance measurements, the observed absorption onset aligns well with the onset energy seen in SE.

While the SE oscillator model reveals the presence of two distinct optical transitions, it does not inherently distinguish whether these TL #2 and TL #3 transitions are direct or indirect. To address this issue, we extract the absorption coefficients $\alpha$ for each transition as shown in Fig. 3(b). The steep absorption onset and the resulting high absorption coefficients for both features indicate that these transitions are direct in nature, as indirect transitions typically exhibit much weaker absorption over a wide spectral range due to phonon-assisted processes. We have attempted to substantiate this statement by examining the data in Tauc plots, but neither $(\alpha E)^{1/2}$ or $(\alpha E)^2$ exhibits an extended linear range suitable to an extrapolation to zero for determining the optical band gap. This finding suggests that the bands involved in the optical transitions are strongly nonparabolic [35].

Our transport measurements have revealed two band gaps, the first at 0.15 eV, which we attribute to the fundamental band gap of CrN, and the second originating from a higher band gap at 0.5 eV [31]. We have no evidence for an optical transition at or close to 0.15 eV, suggesting that the fundamental gap is indirect in nature. Indirect optical transitions are several orders of magnitude weaker than direct ones, and hence cannot be observed by optical techniques in a film of 200 nm thickness. We associate the second electrical gap at 0.5 eV to the direct optical transition with an onset of 0.35 eV. In fact, optical band gaps tend to be smaller than electrical ones due to the existence of tail states [35]. In fact, the absorption coefficient of TL #2 reaches a value characteristic for interband transitions ($> 10^3$ cm$^{-1}$) only at an energy of 0.425 eV. Finally, TL #3 with an absorption onset of 0.6 eV, and an absorption coefficient $> 10^3$ cm$^{-1}$ at 0.67 eV, has no correspondence in our electrical measurements. Considering that the electrical band gap associated with the 0.35 eV absorption onset is detected only at temperatures above 600 K, it is very likely that this even higher energy gap would require temperatures significantly exceeding our maximum one of 920 K.

Figure 4 shows $\varepsilon_1$ and $\varepsilon_2$ of the samples under investigation in the far-infrared range. The spectra of the two samples are essentially identical. Compared to prior reports [5], our films exhibit enhanced amplitudes in both $\varepsilon_1$ and $\varepsilon_2$ due to a lower damping (for a quantitative analysis, see below). The $\varepsilon_1$ spectra cross zero at 50.0 meV (equivalent to 403 cm$^{-1}$ and 12.1 THz) and 78.0 meV (equivalent to 629 cm$^{-1}$ and 18.8 THz), defining the boundaries of the Reststrahlen band [36, 37]. The first zero-crossing corresponds to the transversal optical (TO) phonon mode, marked by the onset of strong absorption and a rise in $\varepsilon_2$. The second zero-crossing is associated with the longitudinal optical (LO) phonon mode. These features are characteristics of infrared-active phonons typically observed in polar materials driven by substantial dynamic charge response and lattice anharmonicity [37]. Notably, the observed phonon energies are comparable to those reported for CrN(001) by Zhang and Gall [5].

In contrast to ScN [38], for which we have observed disorder activated first-order as well as second order Raman-active phonon modes, we were not able to detect any Raman modes from our CrN layers. In our range of excitation energies (1.959 – 3.814 eV), an examination of Fig. 3(b) shows that the combined absorption coefficient for our CrN layers is about $10^6$ cm$^{-1}$, which translates into a Raman information depth of $1/2\alpha \approx 5$ nm. Clearly, this small probe volume is not in favor of strong Raman signals. However, Raman signals were observed in sputtered samples with higher electron densities [5, 39]. Possibly, the elevated carrier density activates scattering mechanisms such as the charge-density fluctuation and the impurity-induced Fröhlich mechanism [40]. In addition, Zhang and Gall [5] report a slightly higher electron density but a significant lower $k$ [Fig. 2(d)], indicating that their layers are less absorptive than our samples. Consequently, this reduced optical absorption may facilitate Raman signal detection by allowing deeper light penetration and reducing re-absorption losses.

To quantitatively analyze the results depicted in Fig. 4, we utilize the expression for the dielectric function of an harmonic oscillator [41]:

$$\varepsilon_L(\omega) = \varepsilon_1(\omega) + i\varepsilon_2(\omega) = \varepsilon_\infty + \frac{(\varepsilon_s - \varepsilon_\infty)\omega_{TO}^2}{\omega_{TO}^2 - \omega^2 - i\gamma_{TO}\omega}, \quad (1)$$

where $\varepsilon_s$ and $\varepsilon_\infty$ denote the relative static (low-frequency) and high-frequency permittivities, respectively. The dielectric function has a pole at $\omega_{TO}$, i.e., the TO phonon frequency, and $\gamma_{TO}$ represents the damping constant of the TO mode. By simultaneously fitting the $\varepsilon_1$ and $\varepsilon_2$ spectra of both films (Fig. 4), we obtain $\varepsilon_s = 39.2 \pm 0.2$ and $\varepsilon_\infty = 15.0 \pm 0.2$, and a damping constant $\gamma_{TO} = 4.90 \pm 0.02$ meV. When compared to the corresponding values reported by Zhang and Gall [5], the TO resonance of our films exhibits a lower damping compared to theirs $6.70 \pm 0.5$ meV, and our values of $\varepsilon_s$ and $\varepsilon_\infty$ are lower than those of Zhang and Gall [5] ($\varepsilon_s \approx 53$, $\varepsilon_\infty \approx 22$). While the differences are not large, they are still significant.

In the absence of free carriers, the Lyddane-Sachs-Teller relation [42] establishes a relation between the permittivities and the optical phonon frequencies of the compound:

$$\frac{\omega_{LO}^2}{\omega_{TO}^2} = \frac{\varepsilon_s}{\varepsilon_\infty}. \quad (2)$$

Using the values derived above, the deviation between both sides is only about 6.9%, which confirms the consistency of our values.

However, this consistency is rather unexpected at the first glance, because our films exhibit an electron density $n_e$ on the order of $10^{19}$ cm$^{-3}$. Because of the coupling of LO phonons with plasmons, we would expect that the LO mode transforms into the high-energy coupled LO phonon-plasmon mode (LPP$^+$) that is notably blueshifted with respect to the LO phonon. With regard to the dielectric

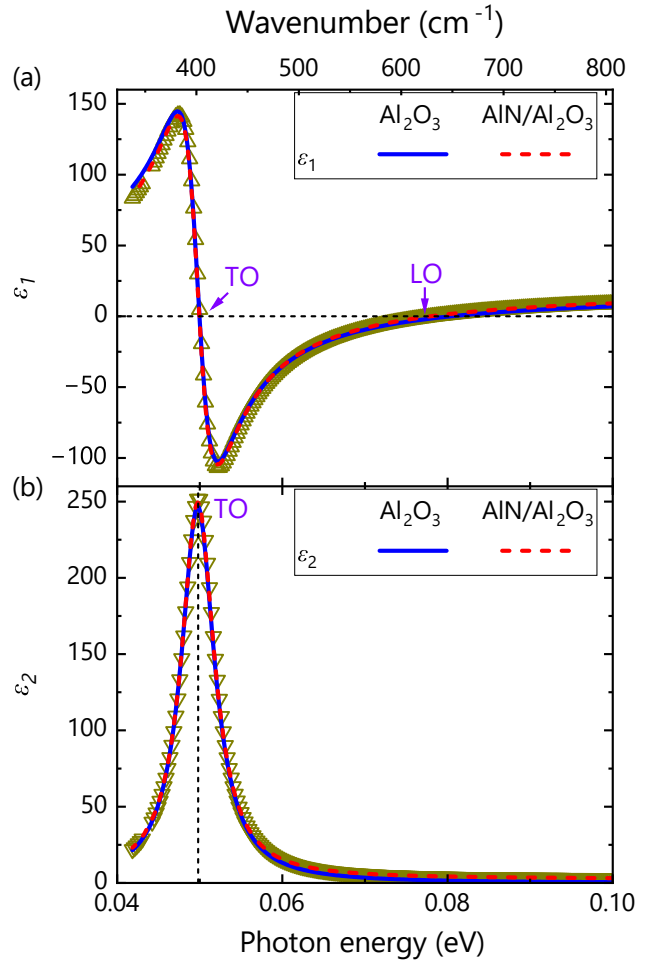

FIG. 4. (a) Real ($\varepsilon_1$) and (b) imaginary ($\varepsilon_2$) components of the complex dielectric function for the CrN thin films in the far-infrared range. Arrows mark the boundaries of the Reststrahlen band, corresponding to TO and LO phonon modes. These experimental data were fitted (symbols) using Eq. 3.

function, this coupling is described by the Drude term:

$$\varepsilon_D(\omega) = -\frac{\varepsilon_\infty \omega_p^2}{\omega^2 - i\gamma_p \omega}, \quad (3)$$

where $\omega_p = e\sqrt{n_e/\varepsilon_s\varepsilon_\infty m_e}$ is the screened plasma frequency with the elementary charge $e$ and the effective electron mass $m_e$, and $\gamma_p = e/m_e\mu_e$ is the plasma damping with the electron mobility $\mu_e$. The influence of this coupling on the LPP$^+$ position and lineshape is of interest because it offers the possibility to optically access $m_e$.

Fits with the total dielectric function $\varepsilon(\omega) = \varepsilon_L(\omega) + \varepsilon_D(\omega)$ return values for $\omega_p$ close to zero, suggesting that the contribution of the Drude term can be neglected. The same behavior was reported by Zhang and Gall [5], who concluded that the electron density in their films must be below the threshold detectable in their infrared spectra. However, simulations show that for electron densities on the order of $10^{19}$ cm$^{-3}$, we would expect a notable influence on the dielectric function provided that $\gamma_p$ remains low.

In fact, LPP$^\pm$ modes with their typical dispersion [43] only form for $\omega_p \gg \gamma_p$. Having independent electrical data of our films, we can estimate these quantities and arrive at $\omega_p < \gamma_p$ for all physically reasonable values of $m_e$ (see also the discussion of the range of electron masses compatible with our transport data in Ref. [31]). In other words, the LPP$^\pm$ excitations in our films are strongly overdamped [40, 44]. In this case, the concept of LPP$^\pm$ modes shifting with plasma frequency does not apply. Regardless of the electron density, we will observe a mode close to the LO phonon energy [40]. The same applies to the previous study of Zhang and Gall [5] considering the electron density and mobility they have published subsequently [6]. Hence, it is not possible to get access to the electron mass in these films via SE unless the electron mobility is enhanced by at least a factor of 10.

Finally, we provide the value of the Born effective charge $Z$ that describes the coupling between macroscopic electric fields and microscopic lattice vibrations [45]. $Z$ quantifies the change in macroscopic polarization induced by a unit displacement of an atomic sublattice within the crystal lattice. Consequently, for materials with a polar bond such as CrN, which exhibit infrared-active TO phonon modes, the Born effective charges of Cr ($Z_{Cr}$) and N ($Z_N$) can be determined from the LO-TO phonon frequency splitting via the relation [46]:

$$\omega_{LO}^2 - \omega_{TO}^2 = \frac{e^2}{\varepsilon_0 \varepsilon_\infty V} \left( \frac{Z_{Cr}^2}{M_{Cr}} + \frac{Z_N^2}{M_N} \right) \quad (4)$$

Here, $V = a^3/4$ denotes the volume of the primitive unit cell of CrN, and $\varepsilon_0$ is the vacuum permittivity. The quantities $M_{Cr}$ and $M_N$ refer to the atomic masses of Cr and N, respectively. Charge neutrality requires $Z_{Cr} = -Z_N$. Using the experimentally determined values for $V$, $\varepsilon_\infty$ and $\omega_{TO,LO}$, the value of the Born effective charge can be derived as $Z_{Cr} = -Z_N = 2.65 \pm 0.01$. When comparing our value to the one of 4.4 reported by Zhang and Gall [5], we note that we obtain a value of 3.1 when using their values for the parameters in Eq. 4. A value of 4.4 is obtained only if we erroneously assume $V$ to be given to $a^3/2$, instead of the correct value of $a^3/4$. Regardless of the reason, the values seem to agree quite well, which calls for comparisons with theoretical values for the paramagnetic phase of CrN.

To summarize and conclude, we have used SE across a wide spectral range (0.04 − 5.5 eV) to extract the optical constants of CrN thin films grown by PAMBE. Spectral line shape fits revealed two distinct optical transitions with onset energies at 0.35 and 0.60 eV, corresponding to higher-energy band-to-band excitations. A pronounced Reststrahlen band from strong TO and LO phonon modes at 403 and 629 cm$^{-1}$ was observed, respectively. The large difference between the static ($\approx$ 39) and high-frequency ($\approx$ 15) dielectric constants, along with a Born effective charge of $\approx$ 2.7, points to a partially ionic bonding character. We also identify an overdamped plasma as the reason for the apparent absence of LO phonon-plasmon coupling despite the high electron density in our films. These findings position CrN as a compelling material platform for infrared-active, thermoelectric, magnetic, and polaritonic devices, where strong light-matter coupling and robust phonon dynamics play a central role.

See supplementary material for (1) 2θ–ω XRD scans and thickness simulation of the films; (2) XRD measurements used for the calculation of lattice constant; (3) Ψ and Δ spectra and their fitting for the CrN/Al$_2$O$_3$ film across the entire spectral range; (4) FTIR transmittance spectra of the CrN/AlN/Al$_2$O$_3$ film.

This work is supported by the Deutsche Forschungsgemeinschaft (DFG, Germany Research Foundation) within the Priority Programme SPP2477 "Nitrides4Future – Novel Materials and Device Concept" through Grant No. 563156864. In addition, we thank Carsten Stemmler for technical assistance with the MBE#9 system. The authors are grateful to Wenshan Chen for a critical reading of the manuscript. We also thank Markus Pristovsek at Nagoya University for supplying AlN templates. This material was based on research sponsored by the Air Force Research Laboratory under Agreement No. FA9453-19-C-1002. The U.S. Government is authorized to reproduce and distribute reprints for Governmental purposes notwithstanding any copyright notation thereon. The views expressed are those of the authors and do not reflect the official guidance or position of the United States Government, the Department of Defense or of the United States Air Force. The appearance of external hyperlinks does not constitute endorsement by the United States Department of Defense (DoD) of the linked websites or the information, products, or services contained therein. The DoD does not exercise any editorial, security, or other control over the information you may find at these locations. Approved for public release; distribution is unlimited.

# Supplementary material: Reststrahlen band and optical bandgaps in semiconducting CrN films

Duc V. Dinh,[*] Xiang Lü, and Oliver Brandt

*Paul-Drude-Institut für Festkörperelektronik, Leibniz-Institut im Forschungsverbund Berlin e.V., Hausvogteiplatz 5–7, 10117 Berlin, Germany.*

Dilara Sen, Olivia Fairlamb, and Frank Peiris

*Department of Physics, Kenyon College, Gambier, 43022 Ohio, United States*

Farihatun Lima, Alexander Bordovalos, Suresh Chaulagain, Ambalanath Shan, and Nikolas J. Podraza

*Department of Physics, University of Toledo, Toledo, 43606 Ohio, United States*

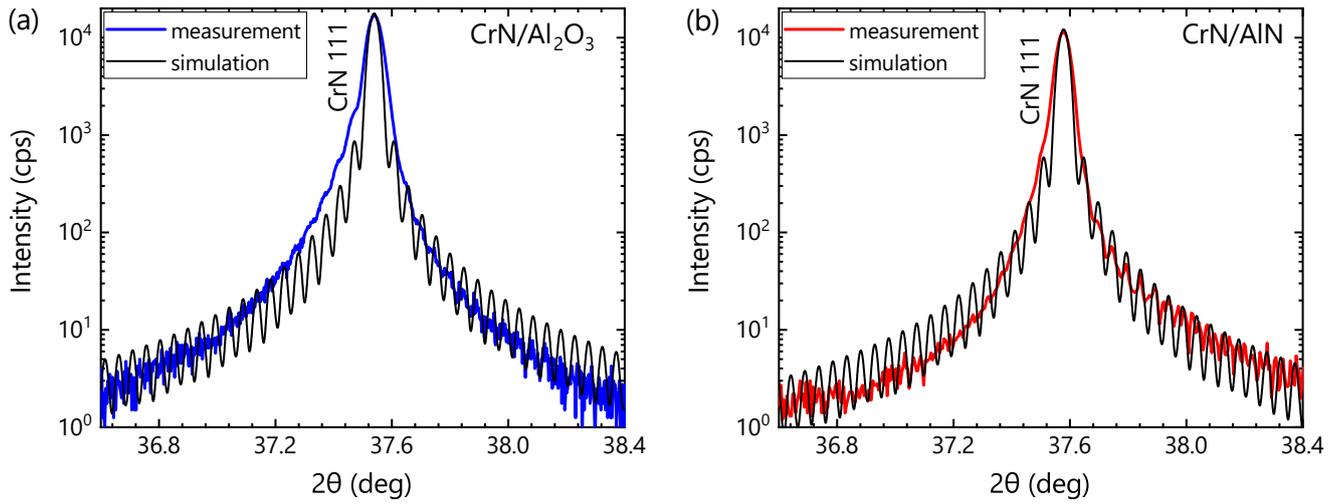

Fig. S1. Experimental and simulated $2\theta-\omega$ XRD scans of the CrN films grown simultaneously on (a) $Al_2O_3$(0001) and (b) $AlN/Al_2O_3$(0001). Both films have similar thickness of $\approx$ 200 nm.

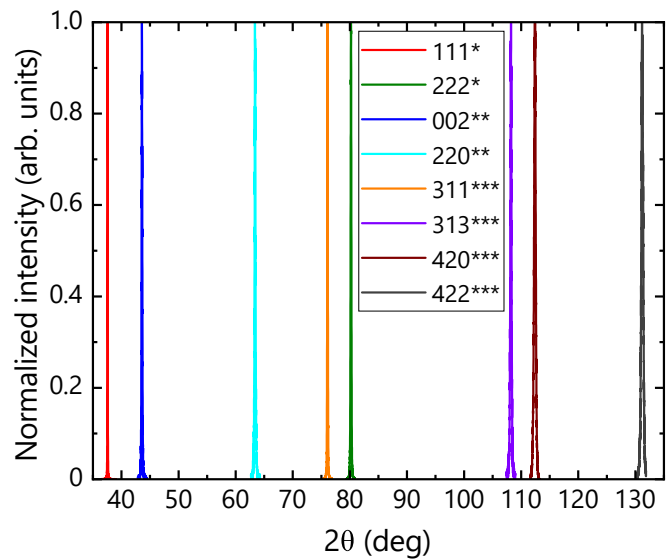

Fig. S2. Symmetric (*), skew-symmetric (**), and asymmetric (***) XRD scans accross the different reflections acquired at room temperature of the CrN(111) film grown on $AlN/Al_2O_3$(0001). From these measurements, we derive the lattice constant of CrN to be $(4.145 \pm 0.001)$ Å.

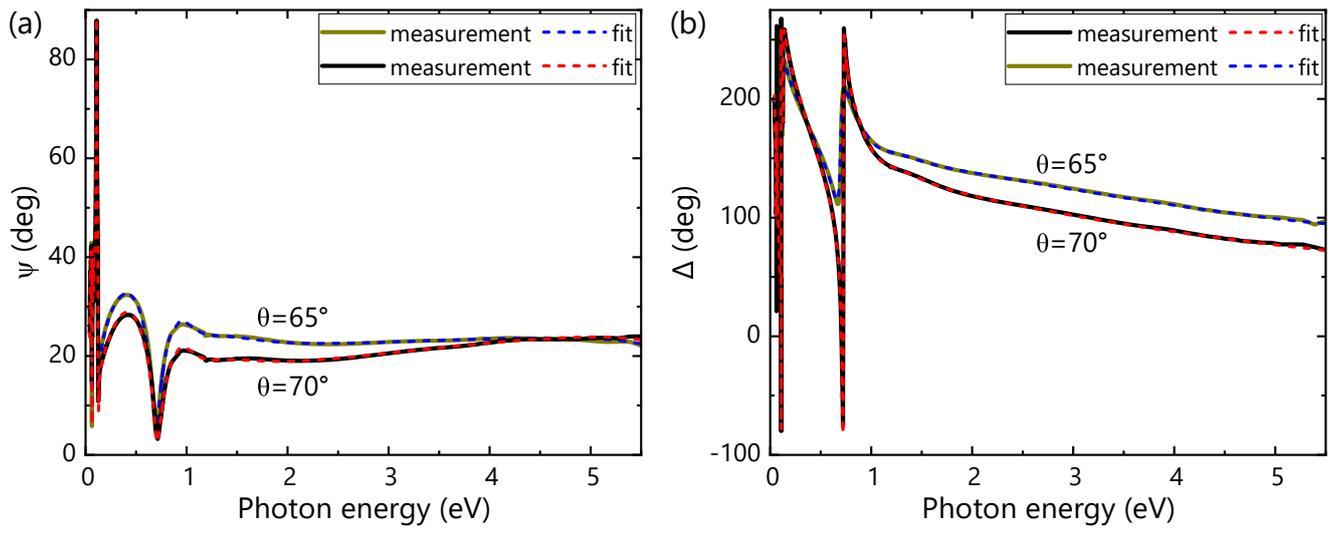

Fig. S3. (a) Ψ and (b) Δ spectra and their fits for the CrN/Al$_2$O$_3$ film.

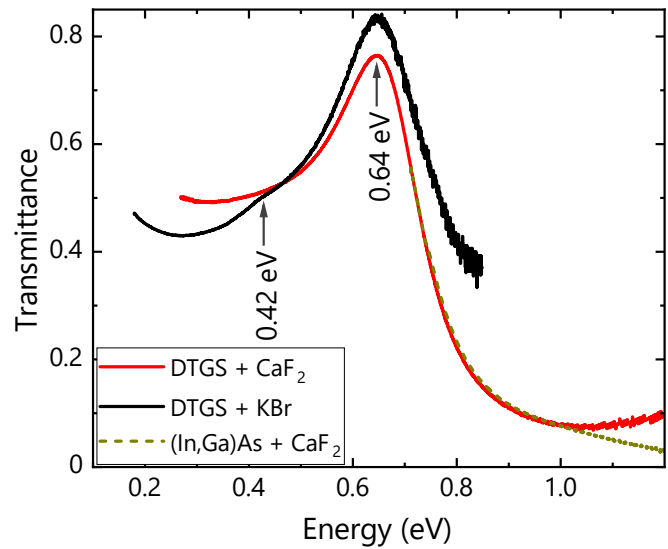

Fig. S4. Room-temperature transmittance spectra of the CrN(111) thin film grown on AlN/Al$_2$O$_3$(0001), measured using three detector–beam splitter configurations: DTGS (Deuterated Triglycine Sulfate) with KBr (mid-IR), DTGS with CaF$_2$ (near-IR), and (In,Ga)As with CaF$_2$ (near-IR). All spectra were normalized to those of the bare AlN/Al$_2$O$_3$(0001) template acquired under identical conditions to ensure proper background subtraction and eliminate the influence of the system response function.

Transmittance measurements were performed on the CrN(111) film grown on AlN/Al$_2$O$_3$(0001) with a Fourier-transform infrared (FTIR) spectrometer (Bruker Vertex 80). A dominant absorption onset is observed at approximately 0.64 eV, in agreement with previous reports [1–3]. This feature is consistent with the intraband transition associated with an absorption onset at 0.60 eV identified from the infrared spectroscopic ellipsometry (IR SE) analysis in the main text. However, because only FTIR transmittance was measured (no reflectance), quantitative extraction of the dielectric function is not possible. The observed peak position does not directly represent the bandgap but instead arises from the combined effects of free carrier absorption and interband transitions. Additionally, the weak shoulder observed at 0.42 eV in the FTIR transmittance spectrum measured in mid-IR is likely associated with the 0.35 eV interband transition identified by SE.